\documentclass[12pt,preprint]{aastex}

\newcommand{\thetahat}{$\hat{\theta}_{mag}$}
\newcommand{\phihat}{$\hat{\phi}_{mag}$}
\newcommand{\psidelay}{$\psi_{delay}$}

\newcommand{\Bvector}{$\vec{B}$}
\newcommand{\Evector}{$\vec{E}$}
\newcommand{\Vvector}{$\vec{v}$}

\newcommand{\adpm}{$a_{1_{DPM}}$}
\newcommand{\avpm}{$a_{0_{VPM}}$}
\newcommand{\anot}{$a_{0_{DPM}}$}
\newcommand{\el}{$\ell$}

\newcommand{\al}{$\alpha$}
\newcommand{\bt}{$\beta$}
\newcommand{\Pone}{$P_1$}
\newcommand{\Ptwo}{$P_2$}
\newcommand{\Pthree}{$P_3$}

\newcommand{\gmodes}{{\it g-}modes}

\shorttitle{Orthogonal Polarization Modes in Radio Pulsars}
\shortauthors{Clemens and Rosen}

\begin{document}

\title{A Pulsational Model for the Orthogonal Polarization Modes in Radio Pulsars}

\author{J. Christopher Clemens \& R. Rosen}
\affil{Department of Physics and Astronomy, University of North 
Carolina, Chapel Hill, NC 27599-3255}
\email{clemens@physics.unc.edu, rrosen@physics.unc.edu}

\begin{abstract} 

In an earlier paper, we introduced a model for pulsars in which non-radial oscillations of high spherical degree (\el) aligned to the magnetic axis of a spinning neutron star were able to reproduce subpulses like those observed in single-pulse measurements of pulsar intensity.  The model did not address polarization, which is an integral part of pulsar emission.  Observations show that many pulsars emit radio waves that appear to be the superposition of two linearly polarized emission modes with orthogonal polarization angles.  In this paper, we extend our model to incorporate linear polarization.  As before, we propose that pulsational displacements of stellar material modulate the pulsar emission, but now we apply this modulation to a linearly-polarized mode of emission, as might be produced by curvature radiation.  We further introduce a second polarization mode, orthogonal to the first, that is modulated by pulsational velocities.  We combine these modes in superposition to model the observed Stokes parameters in radio pulsars.

\end{abstract}

\section{Introduction}
\label{intro2}

In \markcite{cle04}{Clemens} \& {Rosen} (2004), we introduced an oblique pulsator model \markcite{kur82}({Kurtz} 1982) for radio pulsars in which drifting subpulses are reproduced by non-radial oscillations whose periods are incommensurate with the pulsar spin period.  The non-radial modes of our model are aligned to the pulsar magnetic axis, so in addition to the drifting time-like pulses, our model produces longitude stationary variations caused by nodal lines rotating past our line of sight.  Although our model only includes seven parameters, it is able to reproduce a wide variety of observed behavior, including drifting and quasi-stationary subpulses, driftband curvature, and subpulse phase jumps.  It is also able to account for correlations of subpulse phase between the pulse and interpulse of interpulsars as recently discovered in  PSR B1702-19 \markcite{wel07}({Weltevrede}, {Wright}, \&  {Stappers} 2007).

     As we presented it, our model did not attempt to incorporate any polarization effects. This is a major shortcoming; pulsars emit highly polarized radiation with subtle and interesting properties.  In this paper we remedy this shortcoming, using only phenomena associated with non-radial oscillations.  The model of our previous paper \markcite{cle04}({Clemens} \& {Rosen} 2004) then represents a special case of the more general model presented here, and the former maintains its success in reproducing data from a specific subset of pulsars.  However, the more general model presented here allows us to reproduce a wider variety of pulsar behavior, including ``orthogonal polarization modes'' and their interplay with pulse longitude.

     We begin in \S\ref{polarization} with a summary of the (linear) polarization behavior of pulsars.  Following this, in \S\ref{quantities}, we introduce the extensions to our non-radial oscillation model.  These are phenomenological rather than physical, in that they represent the hypothetical effects of non-radial oscillations (in either the star or magnetosphere) without any established model for how the radio emission is actually produced. In \S\ref{qualitative}, we describe the wide range of possible behaviors that result from our model.  We describe detailed analysis and modeling of PSR B0943+10 in a companion paper \markcite{ros07}({Rosen} \& {Clemens} 2007).

\section{Polarization in Pulsars}
\label{polarization}

     The publication by \markcite{rad69}{Radhakrishnan} \& {Cooke} (1969) of their single vector model for pulsar emission was a watershed in the study of radio pulsars, because it convincingly united diverse polarization angle behavior around an intelligible principle.   The linear polarization angle in an average pulsar profile rotates following a vector that points from the site of emission toward the magnetic pole.  That pole is also the epicenter of our non-radial oscillations, and thus we shall see that it is possible to connect our non-radial oscillation model and the rotating vector model into a single elegant and effective description.  

\begin{figure}
\begin{center}
\includegraphics{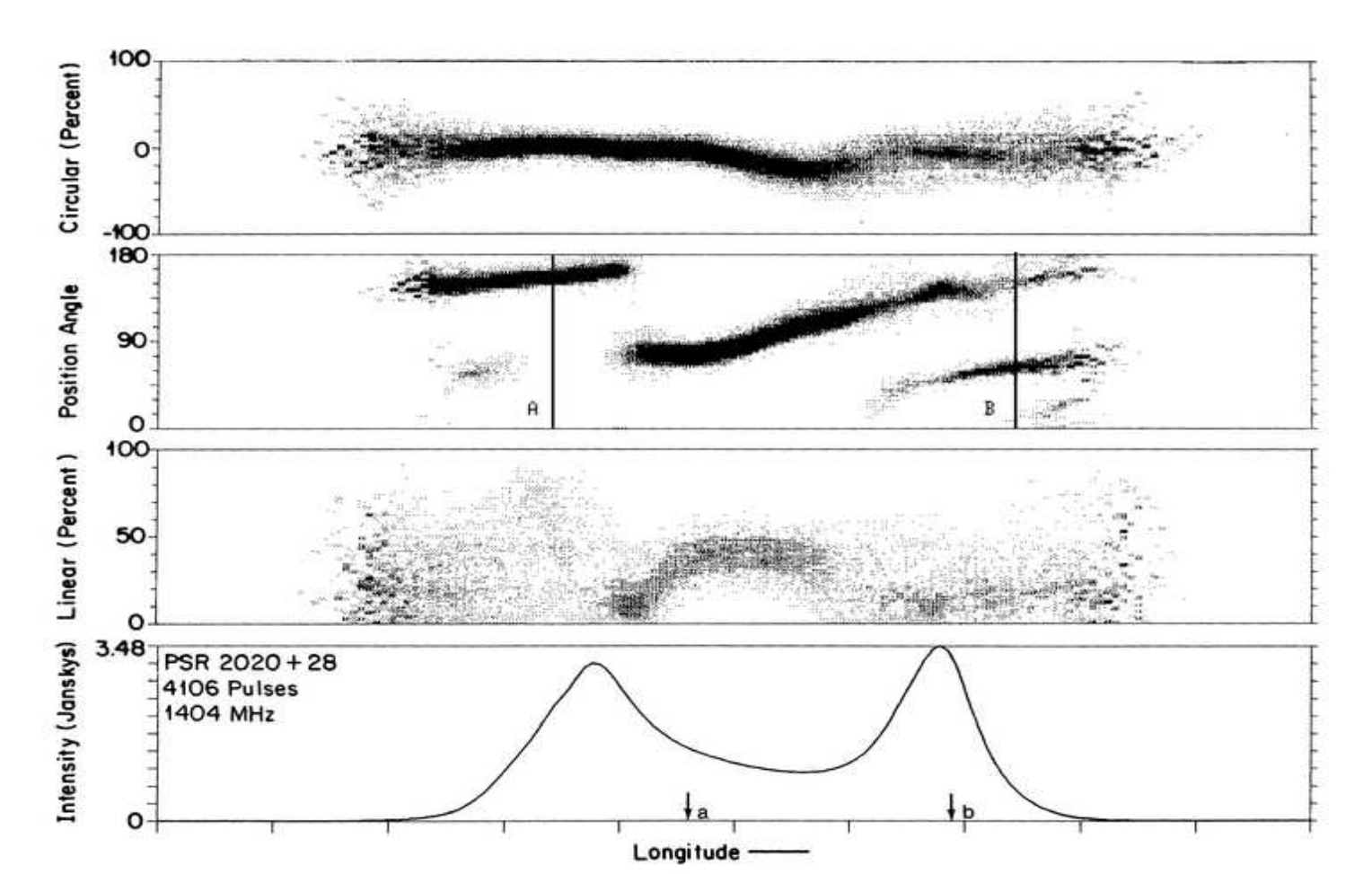}
\end{center}
\caption{Polarization of PSR B2020+28 \markcite{sti84a}({Stinebring} {et~al.} 1984).  The top three panels are histograms showing the circular polarization percentage (top), the linear polarization angle (second), and the linear polarization percentage (third).  The average pulse shape is in the bottom panel.}
\label{fig:sb1984}
\end{figure}

     Once \markcite{rad69}{Radhakrishnan} \& {Cooke} (1969) had accounted for the rotation of linear polarization with pulse longitude, a new complication arose in the form of ``orthogonal polarization modes'' \markcite{bac76a}({Backer}, {Rankin}, \& {Campbell} 1976).  At some pulse longitudes, the polarization angle occasionally jumps by 90 degrees, but continues to follow the rotating vector model on the orthogonal track.  In some pulsars there is more than one switch between the tracks.  The second panel of Figure \ref{fig:sb1984}, reproduced from \markcite{sti84a}{Stinebring} {et~al.} (1984), shows this behavior in the form of a histogram at each pulse longitude.   At each longitude, the histogram counts the number of individual pulses that exhibit a particular polarization angle.  The clear preference for one or the other orthogonally polarized modes is evident, as is the swing in angle associated with the changing vector to the magnetic pole.  In those regions where the two modes occur with nearly equal frequency the linear polarization fraction is reduced (panel three of Figure \ref{fig:sb1984}).  From this behavior, \markcite{sti84a}{Stinebring} {et~al.} (1984) deduced that the orthogonal polarization modes are not disjoint (occurring one at a time but never simultaneously), but rather ``superposed''.  That means the radiation we detect is the superposition of two simultaneously present emissions.  The polarization angle will be that of the higher intensity mode, but if they are exactly equal, there will be complete depolarization.  We will model this situation mathematically, using Stokes parameters for the two modes, in section \S\ref{quantities}.

    Another common property of orthogonally-polarized modes, evident in Figure \ref{fig:sb1984}, is the tendency for the switch between modes to happen repeatedly at fixed pulse longitude.  For instance, the switch in PSR B2020+28 is shown in the second panel in Figure \ref{fig:sb1984}, centered around the regions marked by the lines ``A'' and ``B''.  Because it is longitude-stationary, this feature cannot be associated with rotating structure on the stellar surface \markcite{ran86}({Rankin} 1986) and in many pulsar models reproducing it requires the ad hoc superimposition of unrelated phenomena, e.g. birefringent double imaging of the circulating sparks \markcite{pet00,pet00a}({Petrova} 2000; {Petrova} \& {Lyubarskii} 2000).  In our model, this feature arises from the longitude-stationary pattern of pulsation nodes and antinodes.  

From the scatter at each longitude in the polarization angle histogram, it is evident that the polarization in \textit{individual} pulses can vary from the dominant mode.  This arises because the polarization angle also rotates substantially with the phase of the drifting subpulse.  That is, the rotating vector model describes the behavior of the polarization angle averaged over many pulses.  Within an individual pulse, the polarization may change in correlation with subpulse phase rather than longitude \markcite{man75}({Manchester}, {Taylor}, \&  {Huguenin} 1975).  This polarization angle behavior is analogous to the intensity behavior of individual and average pulse shapes: individual pulses are dominated by the phase of the drifting subpulses, while average pulse shapes show the envelope of their average intensities at each longitude, as we explored in \markcite{cle04}{Clemens} \& {Rosen} (2004).  It is important to recognize that this polarization angle change within individual subpulses can contribute to the depolarization of the average profiles \markcite{cor77}({Cordes} \& {Hankins} 1977), even when all the individual pulses are fully polarized.  
\clearpage
\clearpage

\subsection{\textit{The First Polarization Mode}}
\label{dpm}

     The implication of orthogonal polarization modes, considered in light of the rotating vector model, is that there are two highly linearly-polarized sources of radiation (or one source that is later separated in two, but this has been criticized on sound grounds by \markcite{mic91}{Michel} (1991)).  Almost all of the radio emission mechanisms reviewed by \markcite{mel95}{Melrose} (1995) can produce one linearly polarized component of radiation, but not the other.  Accordingly, in our previous work, we considered only one source of radiation, and proposed that it was modulated by non-radial oscillations of high azimuthal degree (\el).  In this paper, we retain the modulation by pulsational displacements, and explicitly connect it to the radio emission whose transverse electric field vector points toward or away from the magnetic pole, in concordance with the single vector model of \markcite{rad69}{Radhakrishnan} \& {Cooke} (1969) (i.e. in the $\pm$ \thetahat~ direction as defined in Figure \ref{fig:dipole}).  We refer to this radiation as the ``displacement polarization mode'' to remind us of its connection to pulsational displacements.  We express the time-dependent amplitude of this radiation mathematically as the positive portion of the function:

\begin{equation}
\label{eqn:dpm}
A_{DPM}(t) = a_{0_{DPM}} + a_{1_{DPM}} \Psi_{l,m=0}(\theta_{mag})\cos(\omega{t}-\psi_0-\psi_{delay}))
\end{equation}
\\where \anot~ and \adpm~ are constant amplitudes and $\Psi_{l,m=0}$ is a spherical harmonic of $m=0$.  The $\theta_{mag}$ refers to magnetic co-latitude, because the pulsations in our model are aligned to the magnetic pole. This expression is slightly more general than in \markcite{cle04}{Clemens} \& {Rosen} (2004) because it explicitly includes an unmodulated emission baseline (\anot), which we discussed in \markcite{cle04}{Clemens} \& {Rosen} (2004) but, for simplicity, did not incorporate into our mathematical function.  The negative portions of this function, if any, would represent the emission of less than zero light, and for this reason we discard them.  The new expression also breaks the phase into two terms, one of which allows for the arbitrary phase of the drifting subpulses, and the other of which allows for a time lag between the maximum amplitude of the pulsations and emission maximum (\psidelay), which we explain below.  

In \markcite{cle04}{Clemens} \& {Rosen} (2004), we speculated about how pulsational displacements could affect the intensity of the radio emission, but without knowing the radio emission mechanism, or even the site of the pulsations, convincing physical arguments were impossible.   We preferred models in which the pulsations are in the neutron star \markcite{str92,mcd88}({Strohmayer} 1992; {McDermott}, {van Horn}, \&  {Hansen} 1988), rather than in its magnetosphere \markcite{gog05}({Gogoberidze} {et~al.} 2005).  \markcite{str92}{Strohmayer} (1992) proposed that neutron star oscillations could modulate the radio intensity if greater quantities of plasma are injected into the magnetosphere during pulsation maxima, when local heating of the stellar surface is greatest.  This remains a sensible suggestion in light of the conclusion by \markcite{jes01}{Jessner}, {Lesch}, \& {Kunzl} (2001) that thermal emission of electrons dominates the rate at which charged particles flow from the star to magnetosphere.  Following this reasoning, we continue to assume in this paper that the amplitude of the displacement polarization mode, as defined above, follows surface thermal variations caused by non-radial oscillations of the neutron star.  In this respect, the neutron star oscillations we propose are analogous to non-radial oscillations of white dwarf stars, in which the oscillations generate localized heating of the surface material \markcite{rob82}({Robinson}, {Kepler}, \& {Nather} 1982).  This means that for non-adiabatic oscillations, the thermal maximum can lag the displacements in phase. We have included \psidelay~ explicitly to allow for this effect.  

This model is directly analogous to the white dwarf stars, except in white dwarfs the thermal variations directly modulate the radiative flux from the stellar surface while in neutron stars the changing radio flux is a secondary effect of the modulations in plasma emission. The \psidelay~ term is absolutely required in models of white dwarf pulsations \markcite{vank00}({van Kerkwijk}, {Clemens}, \&  {Wu} 2000), and its size has been measured for several pulsation modes in G29-38.  For alternative modulation mechanisms unrelated to surface temperature \markcite{gog05}({Gogoberidze} {et~al.} 2005), \psidelay~ may not be necessary and could then be set to zero.  We show in a companion paper \markcite{ros07}({Rosen} \& {Clemens} 2007) that the sign and magnitude of \psidelay~ are consistent with what we expect from non-adiabatic oscillations.

We explored the properties of a model based on Equation \ref{eqn:dpm} in \markcite{cle04}{Clemens} \& {Rosen} (2004) (for the choice \anot~= 0, $\psi_{delay}=0$), and showed it to be a good model for some pulsars but not others.  In particular it was successful in PSR B1919+21, PSR B1237+25 and PSR B0320+39.  Interestingly these pulsars do not show orthogonal polarization mode switching in the sections of the profiles we modeled, implying that they are dominated by a single polarization mode.  They are therefore exactly the kind of pulsars that should be amenable to modeling with only Equation \ref{eqn:dpm}, a congruence we noticed only after the development of the model in this paper.  

\subsection{\textit{The Second Polarization Mode}}
\label{vpm}

\begin{figure}
\begin{center}
\includegraphics{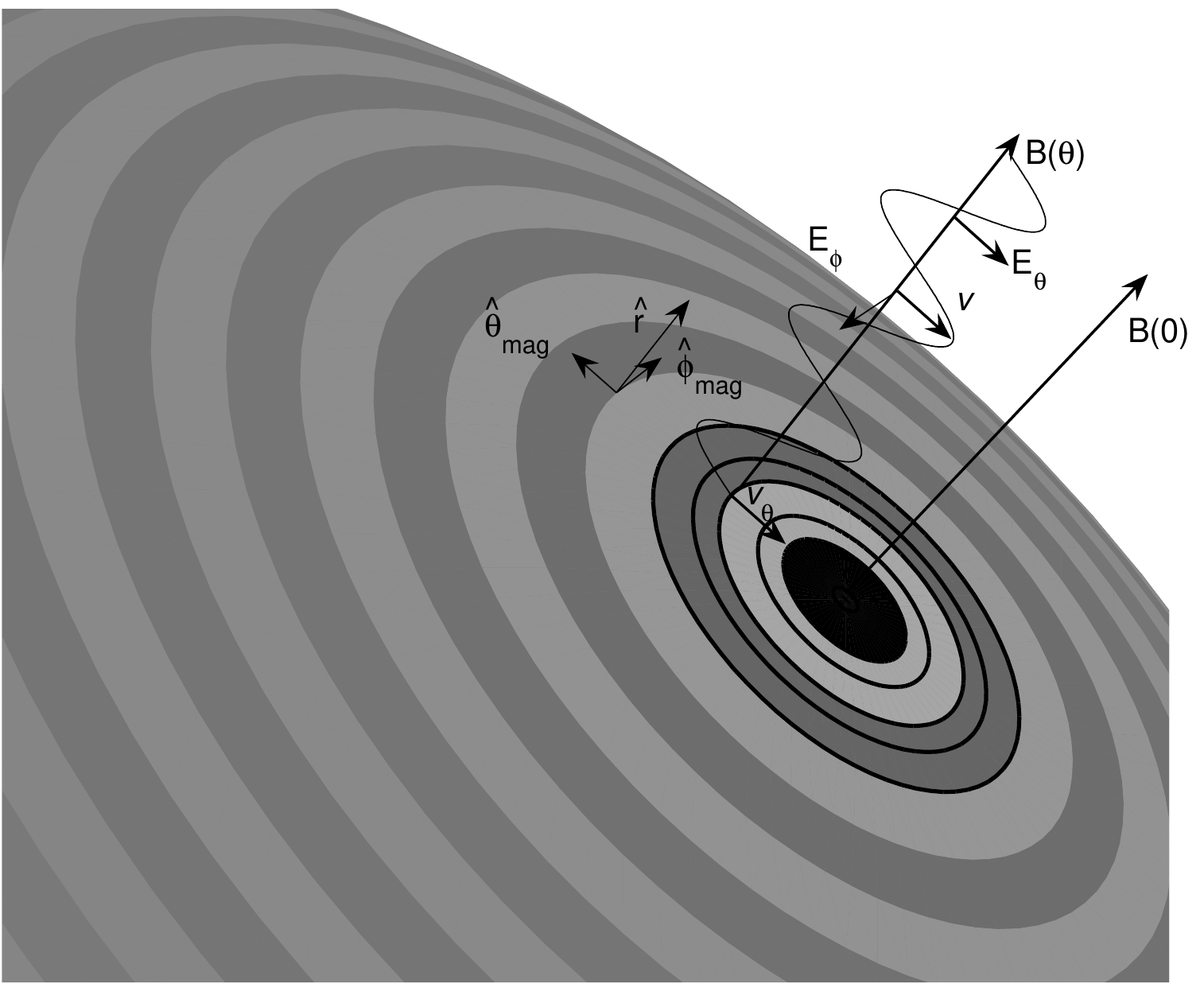}
\end{center}
\caption{The polarization geometry near the surface of a neutron star.  The magnetic field $B(0)$ extends outward, normal to the stellar surface.  The electric field has two components: $E_{\hat{\theta}}$ points in a longitudinal direction and $E_{\hat{\phi}}$ is oriented in a latitudinal direction with respect to the magnetic pole.  The dominant velocity vector \Vvector~ points in the $\pm$ \thetahat~ direction, toward and away from the magnetic pole.  The dark and light areas represent nodal regions of opposing phase.}
\label{fig:dipole}
\end{figure}

In order to model pulsars that emit radiation in two orthogonal polarization modes, we must include a component of radio emission with a transverse electric vector orthogonal to the Radhakrishnan and Cooke vector used in our displacement polarization mode (i.e. in the $\pm$ \phihat~ direction in Figure \ref{fig:dipole}).  Existing models with this property are few.  The maser mechanism of \markcite{fun04}{Fung} \& {Kuijpers} (2004) produces such radiation by the ad hoc imposition of a ``wiggler'' oscillation with transverse \Evector~ vector pointing in the $\pm$ \phihat~ direction (see Figure \ref{fig:dipole}).   The Cherenkov Drift mechanism of \markcite{lyu99}{Lyutikov}, {Blandford}, \&  {Machabeli} (1999a) produces the same polarization in a more natural way, by orienting the vector Cherenkov drift velocity ($u_d$) along the $\pm$ \phihat~ direction.  Figure \ref{fig:dipole} shows this orientation along an imaginary line that represents the magnetic field.  The drift velocity $u_d$ of \markcite{lyu99}{Lyutikov} {et~al.} (1999a) arises from the cyclotron-Cherenkov mechanism operating in a weakly inhomogeneous magnetic field.  
\clearpage
\clearpage

Interestingly, the pulsations we introduced in \markcite{cle04}{Clemens} \& {Rosen} (2004) are all $m=0$ non-radial modes centered on the magnetic pole.  For non-radial \gmodes, the dominant pulsational motions are horizontal, which for $m=0$ modes means the displaced material has a velocity $v_{\theta} \propto \frac{\partial{\Psi}}{{\partial}{\theta}} \frac{\partial{cos(\omega{t})}}{{\partial}t}$ \markcite{dzi77}({Dziembowski} 1977).  Thus, in our oblique pulsator model, the dominant velocity $v_{\theta}$ is directed toward and away from the magnetic pole of the pulsar (Figure \ref{fig:dipole}).  If the surface material interacts with the \Bvector~ field, introducing transverse wiggles as shown in Figure \ref{fig:dipole}, this would generate a transverse electric field modulation,  $\vec{E} =\vec{v}\times\vec{B}$, that is also in the \phihat~ direction.   This suggests that the polarization mode under consideration could be modulated in synchronism with the pulsational velocities, whose magnitudes determine the size of the modulating \Evector~ field.  It also raises the question of whether the pulsations could themselves operate as the wiggler mechanism of \markcite{fun04}{Fung} \& {Kuijpers} (2004).  The subpulses we consider in this paper have frequencies too low ($\propto$ 30 Hz) relative to the radio emission frequencies for this to work, unless there are also unresolved high frequency oscillation modes present.   For the Cherenkov drift emission mechanism, the transverse motions of the \Bvector~ field can modulate the drift velocity if there is an inhomogeneity in the charged particle distribution in the \thetahat~ direction\footnote{The transverse motions of the \Bvector~ field we propose are equivalent to introducing a time dependent $v_r$ into equation 72 of \markcite{lyu99a}{Lyutikov}, {Machabeli}, \&  {Blandford} (1999b).}.  These modulations would lead directly to a modulation in the Cherenkov Drift radiation.  Whatever the mechanism responsible for the second polarization mode, it need not operate at the same altitude above the neutron star surface as the displacement polarization mode.  Difference in altitude might account for the different spectral indices of the orthogonal polarization modes measured by \markcite{smi06}{Smits} {et~al.} (2006).

Following this reasoning, we propose as a hypothesis that non-radial pulsations exactly like those we described in \markcite{cle04}{Clemens} \& {Rosen} (2004) can generate or interact with a second mode of radiation that is linearly polarized in the \phihat~ direction and that the pulsations modulate this radiation by the surface velocities rather than the displacements.  We will refer to this emission as the ``velocity polarization mode''. Mathematically, we model the velocity polarization mode as a time-varying amplitude of the following form:

\begin{equation}
\label{eqn:vpm}
A_{VPM}(t) = a_{0_{VPM}}{{\frac{\partial{\Psi_{l,m=0}}}{\partial{\theta_{mag}}}}}\sin(\omega{t} - \psi_0),
\end{equation}
\\which incorporates the time derivative and the $\theta_{mag}$ derivative of Equation \ref{eqn:dpm}, as appropriate for horizontal pulsational velocities.  This equation is analogous to the $V_{\theta}$ in Equation (3) of \markcite{dzi77}{Dziembowski} (1977).  The $\psi_0$ term is identical to the one in Equation \ref{eqn:dpm}, because it is the phase offset for the same pulsation.  We have dropped the delay term because disturbances in the \Bvector~ field propagate to the emission zone at the Alfv\'{e}n speed, which is near $c$.  \markcite{gil83}{Gil} (1983) have shown the emission altitude to be 10-100 times the neutron star radius, which means that the modulations will be a near-instantaneous representation of the surface velocities.   

Technically, the units of Equation \ref{eqn:vpm} are different from Equation \ref{eqn:dpm}, but we treat both of them as unitless, time-varying amplitudes of the polarized electromagnetic radiation.  The meaning we attach to negative values of Equation \ref{eqn:vpm} is different from that for Equation \ref{eqn:dpm}.  Negative velocities represent a $180^{\circ}$ change in the polarization angle of this mode rather than a reduction of the intensity to values below zero, because they change the sign of $\vec{E} =\vec{v}\times\vec{B}$.  This strategy for interpreting the amplitudes is model dependent; if our reasoning about the emission mechanism is wrong, then a different approach may be required to match the observational data.   We will show in a companion paper that our model generates a satisfactory fit to the observations for PSR B0943+10 \markcite{ros07}({Rosen} \& {Clemens} 2007).

Once more, the situation is analogous to the white dwarf pulsators.  The displacements responsible for surface heating and flux changes in oscillating white dwarfs are primarily horizontal, as discussed by \markcite{rob82}{Robinson}, {Kepler}, \& {Nather} (1982), and can be represented by a formula like our Equation \ref{eqn:dpm}.  More recently, \markcite{vank00}{van Kerkwijk} {et~al.} (2000) have detected the horizontal velocities  associated with these pulsations via radial velocity variations. These variations arise from the horizontal surface motions viewed at the limb of the star \markcite{cle00}({Clemens}, {van Kerkwijk}, \& {Wu} 2000).  As expected, they cause spectral line shifts that follow Equation \ref{eqn:vpm}, although we can only see an integral over the observed hemisphere of the star.

\section{Observed Quantities}
\label{quantities}

To convert the amplitudes in Equations \ref{eqn:dpm} and \ref{eqn:vpm} into observable quantities, we use the following transformations to calculate Stokes parameters in the frame of the star:

\begin{equation}
\label{eqn:I}
I = <A_{DPM}>^2 + <A_{VPM}>^2
\end{equation}

\begin{equation}
\label{eqn:Q}
Q' = <A_{DPM}>^2 - <A_{VPM}>^2
\end{equation} 

\begin{equation}
\label{eqn:U}
U' = 0
\end{equation}

This is equivalent to assuming that the orthogonal polarization modes are completely linearly polarized, and that their superposition generates the emission we observe.  When one or the other mode dominates, the fractional linear polarization is high, and the polarization angle follows the dominant mode; when the two modes have equal amplitudes, complete depolarization occurs.  We have not included circular polarization in the model presented in this paper.

The choice of prime notation for $Q'$ and $U'$ follows the usage of \markcite{des01}{Deshpande} \& {Rankin} (2001), who use primed coordinates to refer to observed orthogonal polarization modes with the rotating vector model removed (i.e. converted to the non-rotating frame). Translating from the primed quantities into the observer's frame requires incorporating the changing longitude we observe as the star spins and imposing rotation of the polarization angle so that it follows the magnetic pole, as we present in our companion paper \markcite{ros07}({Rosen} \& {Clemens} 2007).

\section{Qualitative Behavior of the Model}
\label{qualitative}

\begin{figure}
\begin{center}
\includegraphics[width=150mm,height=80mm]{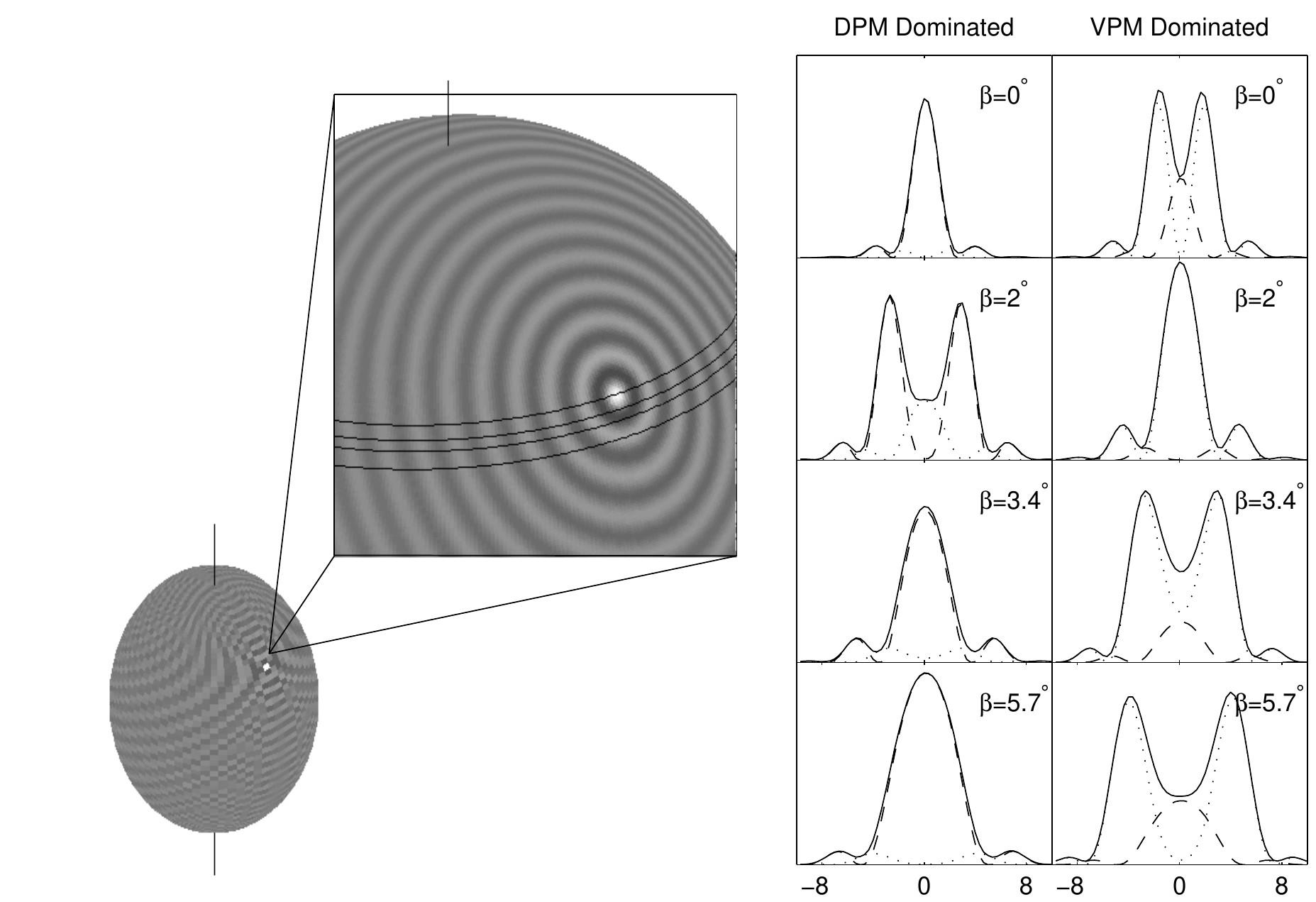}
\end{center}
\caption{The variety of observed average pulses as a function of changing \bt~ created from the two orthogonal polarization modes.  In the panels, the solid line represents the total intensity and the dashed and dotted lines represent the square of the displacement (Equation \ref{eqn:dpm}) and velocity (Equation \ref{eqn:vpm}) polarization modes, respectively.  The panels on the left show a model dominated by the displacement polarization mode and the panels on the right show a model dominated by the velocity polarization mode.  Both models have $\alpha = 50^{\circ}$ and $l=70$.}
\label{fig:sightlines2}
\end{figure}

In this section we explore, briefly, the generic behavior of a model based on Equations \ref{eqn:dpm} and \ref{eqn:vpm}.  Detailed comparison of the model to data are presented in a companion paper \markcite{ros07}({Rosen} \& {Clemens} 2007).  Figure \ref{fig:sightlines2} is an extension of Figure 2 from \markcite{cle04}{Clemens} \& {Rosen} (2004), and shows how the variety of observed average pulse shapes can be generated by changing the viewing geometry (impact parameter \bt).  The panels in Figure \ref{fig:sightlines2} show the averages of the square of the displacement and velocity polarization modes (generated using Equations \ref{eqn:dpm} and \ref{eqn:vpm}), along with the average intensity.  The panels on the left show models dominated by the displacement polarization mode.  The pulse shapes are similar to those in \markcite{cle04}{Clemens} \& {Rosen} (2004), and the subpulse phase changes by $180^{\circ}$ at nodal lines, as in our previous paper.  There are notable differences in our presentation of the model in Figure \ref{fig:sightlines2}.  In our previous work, we plotted displacements directly, now we convert them into Stokes parameters, which are proportional to the square of the amplitudes.  We have added the window function that suppresses intensity at the edges of the profile.  For example, a sightline that crosses directly through the magnetic pole (\bt = $0^{\circ}$) might give an average profile with five components, like the top panel in Figure 2 of \markcite{cle04}{Clemens} \& {Rosen} (2004) or as in the fit to the average profile of PSR B1237+25 in Figure 8 of \markcite{cle04}{Clemens} \& {Rosen} (2004).  The imposition of a window function suppresses the outer components in Figure \ref{fig:sightlines2} so that the top left panel appears to have fewer components.  Previously we had simply truncated the model at the edges of the profile.  

The panels on the right in Figure \ref{fig:sightlines2} show models dominated by the velocity polarization mode.  The maxima in their average pulse shapes occur at nodes instead of antinodes, i.e., the spatial phase is shifted by $90^{\circ}$ because of the derivative in Equation \ref{eqn:vpm}.  Other than the ratio \adpm/\avpm, the panels on the right and left have the same geometric parameters in the model (\al, \bt, $l$).  All of the differences arise from changing which of the two orthogonal modes dominates the profile.  
\clearpage
\clearpage

The drifting subpulse behavior is much richer in this model.  Whereas a displacement-polarization-mode-dominated model can show phase shifts of only $180^{\circ}$, and these only at nodal lines, a model with both polarization modes has subpulse phases that also depend upon the longitude dependent amplitude ratio of the two polarization modes (\adpm, \avpm), and their fixed phase difference (\psidelay).  This means that the subpulse phase can have a larger variety of longitude dependent changes.  Our model predicts that these will be most pronounced in pulsars that show orthogonal mode switching.   These effects can also combine with the phase shifts at nodal lines to result in phase changes that are not immediately equal to 180 degrees.   

\begin{figure}
\begin{center}
\includegraphics[scale=.8]{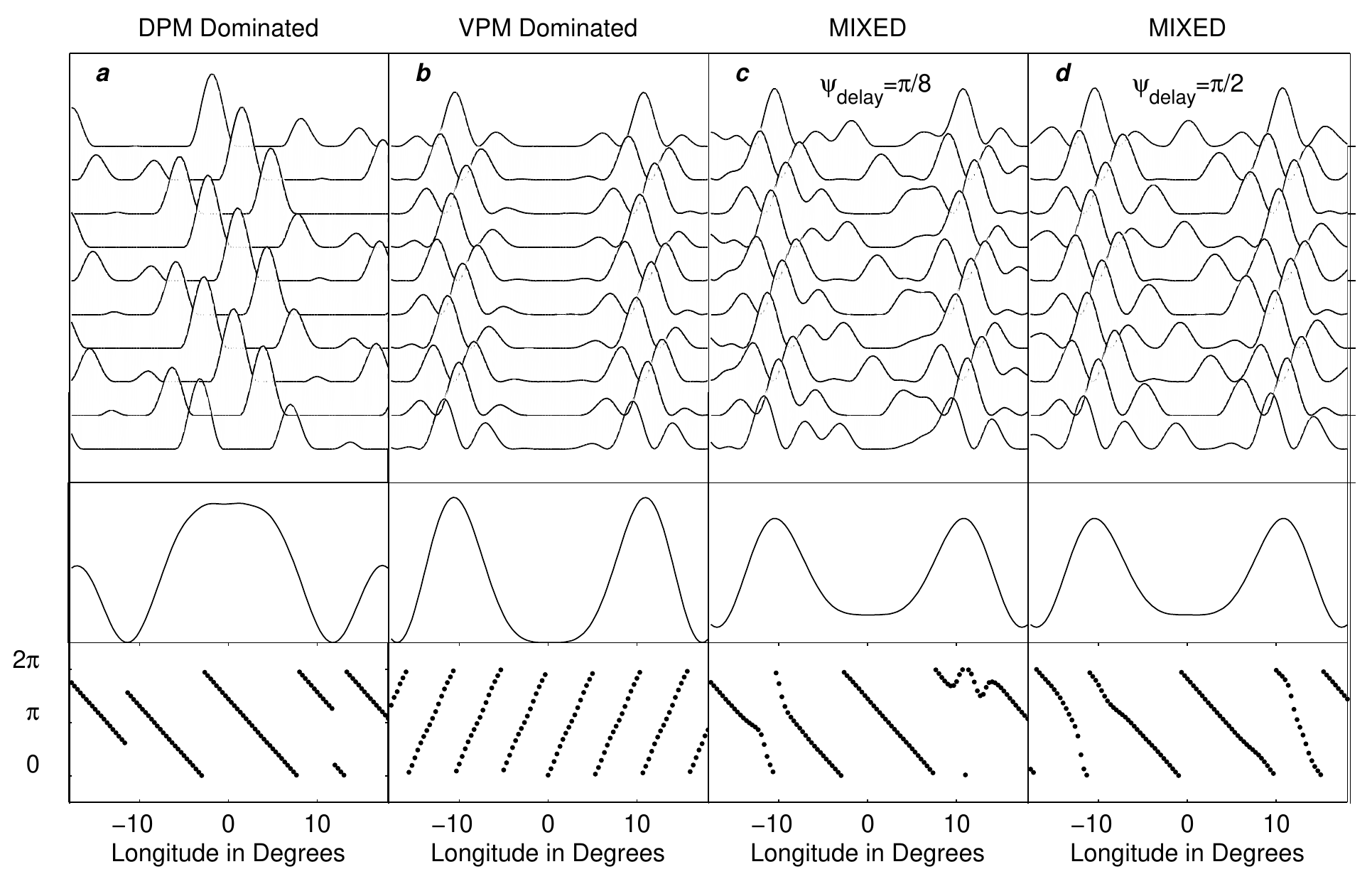}
\end{center}
\caption{Individual subpulses for a sightline traverse with a \bt~ = $-1.7^{\circ}$ (left to right): a displacement polarization mode dominated model, a velocity polarization mode dominated model, a model with both polarization modes present.  The average of 600 pulses for each model are shown in the bottom panels.}
\label{fig:intuition}
\end{figure}

Figure \ref{fig:intuition} shows subpulse behavior for the two polarization modes independently and in combination.  Panel \textit{a} shows a model dominated by the displacement polarization mode, the same case we considered in \markcite{cle04}{Clemens} \& {Rosen} (2004).  The individual pulses are at the top, the average pulse shape is in the middle, and the subpulse phases are at the bottom.  As \markcite{edw06}{Edwards} (2006) has noted, the subpulse phase in this case is a strictly linear function of longitude except where nodal lines introduce a $180^\circ$ phase jump.  The amplitude windowing effects discussed in \markcite{cle04}{Clemens} \& {Rosen} (2004) only affect times of maxima and not phases from sinusoidal fits to the data.  The phases we show were measured by fitting a sinusoid with a period of \Pthree~ at each longitude.  \markcite{edw06}{Edwards} (2006) has correctly understood and described the phase behavior of our model as presented in \markcite{cle04}{Clemens} \& {Rosen} (2004); the displacement polarization mode presented in panel \textit{a} behaves the same way. 

Panel \textit{b} shows the behavior of a model including only our velocity polarization mode, which has the same period but different phase from the displacement polarization mode.  It also has opposite polarization and a different spatial modulation, i.e. its maxima occur at nodal lines and its minima at antinodal lines.  As the phase plot for this mode shows, it appears to have twice the frequency, because its amplitude is the square of a periodic function whose negative portions are not removed.  For the same reason, it does not show $180^{\circ}$ phase changes at antinodes. Its phase is also a strictly linear function of longitude.  The completely different slope in the phase plot arises from different aliasing of this signal with \Pone; its aliasing is that of the first harmonic of \Ptwo~ rather than \Ptwo~ itself.

Panels \textit{c} and \textit{d} show the superposition of the displacement and velocity polarization modes as might be used in a model that reproduces orthogonal mode switching.  Panel \textit{c} is a combination of the two polarization modes for an arbitrary choice of \psidelay~= $\pi/8$.   The model has only one single, stable subpulse frequency, but the drifting subpulse behavior, and the phase patterns are complex and evocative of observed pulsars.  The pulse shape and the subpulse phase behavior depend upon the ratio of the amplitudes of the two polarization modes, and upon their relative phases.  The model in panel \textit{d} is the same as panel \textit{c} except that now \psidelay~= $\pi/2$.  Without detailed pulsational models, we do not know what value to expect for \psidelay~ so this value was chosen to show how subpulses might change with \psidelay.  

This model can account for many observed phenomena not explained by \markcite{cle04}{Clemens} \& {Rosen} (2004).  For example, subpulse phase jumps that occur in only one of the polarization modes \markcite{edw06}({Edwards} 2006), subpulse phase changes that are not instantly $180^\circ$ \markcite{edw03}({Edwards} \& {Stappers} 2003), and apparent changes in driftband slope \markcite{esa05}({Esamdin} {et~al.} 2005). This model conforms in general with the observational description of subpulses in PSR B0809+74 as ``the out-of-phase superposition of two orthogonally polarized drift patterns'' \markcite{edw04,ran03}({Edwards} 2004; {Rankin} \& {Ramachandran} 2003).  Whether it can reproduce even more confused subpulse polarization behavior (e.g. PSR B0818-13, \markcite{edw04}({Edwards} 2004)) is an exercise that will require careful numerical modeling like that we have done for PSR B0943+10 in our companion paper.

\clearpage
\clearpage

\section{Conclusions}
\label{conc2}

In our first paper, \markcite{cle04}{Clemens} \& {Rosen} (2004), we introduced a model for pulsars in which non-radial oscillations of high spherical degree (\el) were aligned to the magnetic axis of a spinning neutron star. The rotation of the pulsar carried a pattern of pulsation nodes underneath our sightline, reproducing the longitude stationary structure seen in average pulse profiles. The associated time-like oscillations reproduced the drifting subpulses. The presence of nodal lines accounted for observed $180^{\circ}$ phase jumps in drifting subpulses and apparent changes in \Ptwo, even when the time-like oscillations were strictly periodic. Our model also accounted for the mode changes and nulls observed in some pulsars as quasi-periodic changes between pulsation modes of different (\el) or radial overtone ($n$), analogous to pulsation mode changes observed in oscillating white dwarf stars.

In this paper we extended our model to account for the diverse polarization behavior observed in pulsars. The measured polarization angles in many pulsars show the existence of orthogonally polarized components of radio emission.  We hypothesized that our initial model described one linear polarization mode, which we call the displacement mode because its variations follow the pulsational displacements.  We then introduced a second mode whose variations follow the pulsational velocities.  We further hypothesized that this mode has a linear polarization orthogonal to the first mode.  The physical basis for this hypothesis is the orthogonal direction of the electric vector generated by the cross product of the magnetic field and the pulsational velocities.   We allowed for non-adiabatic pulsations by the incorporation of a \psidelay~ phase adjustment between the time-like variations of the two modes.

In the absence of a settled model for pulsar emission, the physical basis for our hypotheses is tenuous.  Nonetheless, the results are encouraging; our model not only reproduces the observed polarization behavior, but also generates complex drifting subpulse behavior similar to what we observe.  The variety of subpulse phase changes allowed in our model is greater than before; a necessary step toward modeling pulsars like PSR B0809+74.  

In a companion paper to this one, we fulfill an earlier promise to produce a quantitative model for PSR B0943+10. We have been able to fit data from that star using our model and to generate synthetic pulsar data using the fitted parameters.  The synthetic data capture the essence of the variations in that star better than previous models.  The elements still missing are circularly polarized and unpolarized components of the emission. We are hopeful that the application of our model to a larger sample of pulsars will illuminate the path to understanding these phenomena. 

We believe that our model offers a viable alternative to the drifting spark model.  It is physically plausible, economical, and readily disprovable. Following \markcite{kri80}{Krishnamohan} (1980) and \markcite{wri81}{Wright} (1981), \markcite{edw02}{Edwards} \& {Stappers} (2002) and \markcite{edw06}{Edwards} (2006) have pointed out that the drifting spark model makes specific, testable predictions about subpulse phase. The same is true of our model, but the predictions are different.  Consequently, subpulse phase measurements offer an opportunity to discriminate between the two models.   However, the phase measurements required are subject to distortion by noise and even the naturally varying pulse amplitudes, so the tests require great care.  If our pulsation model survives rigorous observational testing, it will lead directly to physical insight into the radio emission mechanism for pulsars and, more importantly, the physical structure of neutron stars.  

\bibliography{}
\end{document}